\documentclass[11pt]{article}
\usepackage{latexsym}
\usepackage{amsfonts}

\newtheorem{theorem}{Theorem}

\begin{document}
\title{The Einstein-Vlasov system/Kinetic theory} 
\author{H\aa kan Andr\'{e}asson\\
         Department of Mathematics\\
        Chalmers University of Technology\\
        S-412\,96 G\"oteborg,
        Sweden\\hand@math.chalmers.se\\http://www.math.chalmers.se/\~{}
        hand} 
\maketitle

\begin{abstract}

\end{abstract} 
The main purpose of this article is to guide the reader to theorems on 
global properties of solutions to the Einstein-Vlasov system. This
system couples Einstein's equations to a kinetic matter model. Kinetic 
theory has been an important field of research during several decades
where the main focus has been on nonrelativistic- and special
relativistic physics, e.g. to 
model the dynamics of neutral gases, plasmas and Newtonian
self-gravitating systems. In 1990 Rendall and Rein initiated a
mathematical study of the Einstein-Vlasov system. 
Since then many theorems on global properties of solutions to this 
system have been established. The Vlasov equation describes matter 
phenomenologically and it should be stressed that most of the theorems
presented in this article are not presently known for other such
matter models (e.g. fluid models). The first part of this paper gives
an introduction to 
kinetic theory in non-curved spacetimes and then the Einstein-Vlasov
system is introduced. We believe that a good understanding of kinetic
theory in non-curved spacetimes is fundamental in order to get a good 
comprehension of kinetic theory in general relativity. 
\section{Introduction to kinetic theory} 
In general relativity kinetic theory has been used
relatively sparsely to model phenomenological matter in comparison to 
fluid models. From a mathematical point of view there are however
fundamental advantages using a kinetic description. In 
non-curved spacetimes kinetic theory has been studied intensively as a
mathematical subject during several decades and it has also played an
important role from an engineering point of view. In the first part of
this introduction we will review kinetic theory in non-curved 
spacetimes and we will mainly consider the special relativistic case
but mathematical results in the nonrelativistic case
will also be discussed. We believe that 
a good understanding of kinetic theory in non-curved spacetimes is
fundamental in order to 
get a good comprehension of kinetic theory in general relativity. 
Moreover, it is often the case that mathematical methods used to treat
the Einstein-Vlasov system are carried over from methods developed in
the special relativistic- or nonrelativistic case. 

The purpose of kinetic theory is to model the time evolution of a
collection of particles. The particles may be entirely different 
objects depending on the physical situation. For instance, 
the particles are 
atoms and molecules in a neutral gas or electrons and ions in a 
plasma. In stellar dynamics the particles are stars and in a 
cosmological case they are galaxies or even clusters 
of galaxies. 
Mathematical models of particle systems are most frequently described 
by kinetic or fluid equations. 
A characteristic feature of kinetic theory is 
that its 
models are statistical and the particle systems are described 
by distribution functions $f=f(t,x,p)$, which 
represent the density of particles with given space-time position 
$(t,x)\in\mathbb{R}\times\mathbb{R}^3$ and 
momentum $p\in \mathbb{R}^{3}$. A distribution function 
contains a wealth of information, and macroscopical 
quantities are easily calculated from this function. 
In a fluid model the quantities that describe the
system do not depend on the momentum $p$ but only on 
the spacetime 
point $(t,x)$. A choice of
model is usually made with regard to the physical properties of
interest for the system or with regard to numerical considerations. 
It should be mentioned that a too naive fluid model may give rise to
shell-crossing singularities which are unphysical. 
In a kinetic description such phenomena are ruled out. 

The time evolution of the system is determined by the interactions 
between the particles which depend on the physical situation. For
instance, the driving mechanism for the time evolution of a neutral
gas is the collisions between the particles (the relativistic
Boltzmann equation). For a plasma the interaction is 
through the electric charges (the Vlasov-Maxwell system) 
and in the stellar- and 
cosmological case the interaction is gravitational (the
Einstein-Vlasov system). Of course, combinations of interaction
processes are also considered but in many situations one of them is
strongly dominating and the weaker processes are neglected. 

\subsection{The relativistic Boltzmann equation}
Consider a collection of neutral particles in Minkowski spacetime. Let 
the signature of the metric be $(-,+,+,+)$, let all the particles 
have rest mass $m=1$ and normalize the speed of light, $c$, to one. 
The four-momentum of a particle is denoted by 
$p^{a}$, $a=0,1,2,3$. 
Since all particles have equal rest mass, the four-momentum for
each particle 
is restricted to the mass shell, 
$p^{a}p_{a}=-m^{2}=-1$. Thus, by denoting the three-momentum 
by $p\in\mathbb{R}^{3}$, $p^{a}$ may be written 
$p^{a}=(-p^0,p)$, where $|p|$ is the usual 
Euclidean length of $p$ and $p^{0}=\sqrt{1+|p|^{2}}$ is the energy of
a particle with three-momentum $p$. The relativistic 
velocity of a particle with momentum $p$ is denoted by $\hat{p}$ and
is given by 
\begin{equation}
\hat{p}=\frac{p}{\sqrt{1+|p|^{2}}}.\label{velo}
\end{equation}
Note that $|\hat{p}|<1=c$. 
The relativistic Boltzmann equation models the space-time behaviour 
of the one-particle distribution function $f=f(t,x,p),$ and it 
has the form 
\begin{equation}
(\partial _{t} + \frac{p}{p^{0}}\cdot\nabla _{x})f=Q(f,f), \label{rbe}
\end{equation}
where the relativistic collision operator $Q(f,g)$ is defined by 
\begin{eqnarray}
&\displaystyle Q(f,g)=\int_{\mathbb{R}^{3}}\int_{\mathbb{S}^{2}} k(p,q,\omega)\times
&\nonumber\\ &\displaystyle 
\times [f(p+a(p,q,\omega)\omega) g(q-a(p,q,\omega)\omega)-f(p)g(q)] dpd\omega. &
\end{eqnarray}
(Note that $g=f$ in (\ref{rbe})). 
Here $d\omega$ is the element of surface area on $\mathbb{S}^{2}$ and 
$k(p,q,\omega)$ is the scattering kernel which depends on 
the scattering cross-section in the interaction process. See [GLW] for
a discussion about the scattering kernel. The function 
$a(p,q,\omega)$ results from the collision mechanics. 
If two particles with momentum $p$ and $q$ respectively, collide 
elastically (no energy loss) with scattering 
angle $\omega\in\mathbb{S}^{2}$, their momenta will 
change, $p\rightarrow p'$ and $q\rightarrow q'$. 
The relation between $p,q$ and $p',q'$ is 
\begin{equation}
p'=p+a(p,q,\omega)\omega,\,\,\,q'=q-a(p,q,\omega)\omega,
\end{equation}
where 
\begin{equation}
a(p,q,\omega)=\frac{2(p^{0}+q^{0})p^{0}
q^{0}(\omega\cdot(\hat{q}-\hat{p}))}
{(p^{0}+q^{0})^{2}-(\omega\cdot(p+q))^{2}}.
\end{equation}
This relation is a consequence of four-momentum conservation, 
$$p^{a}+q^{a}=p^{a'}+q^{a'},$$ or equivalently 
\begin{eqnarray}
p^{0}+q^{0}&=&p^{0'}+q^{0'},\\ 
p+q&=&p'+q'.
\end{eqnarray}
These are the conservation equations for relativistic particle
dynamics. In the classical case 
these equations read 
\begin{eqnarray}
|p|^{2}+|q|^{2}&=&|p'|^{2}+|q'|^{2},\label{cocl1}\\
p+q&=&p'+q'\label{cocl2}. 
\end{eqnarray}
The function $a(p,q,\omega)$ is the distance between 
$p$ and $p'$ ($q$ and $q'$) and the analogue function 
in the Newtonian case has the form 
\begin{equation}
a_{cl}(p,q,\omega)=\omega\cdot (q-p). 
\end{equation}
By inserting $a_{cl}$ in place of $a$ in (\ref{rbe}) we obtain 
the classical Boltzmann collision operator (disregarding 
the scattering kernel which also is different). 

The main result concerning existence of solutions to the classical Boltzmann
equation is a theorem by DiPerna and Lions [DL1] which proves 
existence, but not uniqueness, of renormalized solutions 
(i.e. solutions in a weak sense which are even more 
general than distributional solutions). An analogous result holds in
the relativistic case which was shown by Dudy\'{n}sky and
Ekiel-Jezewska [DE]. Regarding classical solutions, 
Illner and Shinbrot [IlS] have shown global existence of solutions to
the nonrelativistic Boltzmann equation for small initial data (close to
vacuum). At present there is no analogous result for the relativistic 
Boltzmann equation and this must be regarded as an interesting open
problem. 
If the data are 
close to equilibrium (see below) global existence of classical
solutions has been proved by Glassey and Strauss [GSt3] in the relativistic 
case and by Ukai [U] in the nonrelativistic case (see also [Sh]). 

The collision operator $Q(f,g)$ may be written in an obvious way as 
$$Q(f,g)=Q^{+}(f,g)-Q^{-}(f,g),$$ where $Q^{+}$ and $Q^{-}$ are
called the gain and loss term respectively. 
In [An1] it is proved that given 
$f\in L^{2}(\mathbb{R}^{3})$ and $g\in L^{1}(\mathbb{R}^{3})$ with $f,g\geq 0$, then 
\begin{equation}
\| Q^{+}(f,g)\|_{H^{1}(\mathbb{R}^{3})}\leq C\| f\|_{L^{2}
(\mathbb{R}^{3})}\|g\|_{L^{1}(\mathbb{R}^{3})},\label{Q} 
\end{equation}
under some technical requirements on the scattering kernel. 
Here $H^{s}$ is the usual Sobolev space. 
This regularizing result was first proved by 
P.L. Lions [L] in the classical situation. 
The proof relies on the theory of Fourier integral
operators and on the method of stationary phase, and requires 
a careful analysis of the collision geometry which is very different 
in the relativistic case. 

The regularizing theorem has many applications. An important
application is to prove that solutions tend to equilibrium for large
times. More precisely, Lions used the regularizing theorem
to prove that solutions to the (classical) Boltzmann 
equation with periodic boundary conditions, converge in $L^{1}$ to 
a global Maxwellian, $$M=e^{-\alpha |p|^{2}+\beta\cdot p+\gamma}\;\;
\alpha,\gamma\in R,\;\alpha>0,\;\beta\in \mathbb{R}^{3},$$ 
as time goes to infinity. This result had first been obtained by
Arkeryd [Ar] by using non-standard analysis. It should be pointed out
that the convergence takes place through a sequence of times tending
to infinity and it is not known if the limit is unique or depends on
the sequence. 
In the relativistic situation the analogous question of 
convergence to
a relativistic Maxwellian, or a J\"{u}ttner 
equilibrium solution, 
$$J=e^{-\alpha\sqrt{1+|p|^{2}}+\beta\cdot
  p+\gamma},\;\;\alpha,\beta\mbox{ and }\gamma\mbox
  { as above, with } \alpha>|\beta|,$$ 
had been studied by Glassey and Strauss [GSt3], [GSt4]. 
In the periodic
case they proved 
convergence in a variety of function spaces for initial data close to
a J\"{u}ttner solution. 
Having obtained the regularizing theorem for the relativistic 
gain term, it is a straightforward task to follow the method of Lions
and prove convergence to 
a \textit{local} J\"{u}ttner solution for 
arbitrary data (satisfying the natural
bounds of finite energy and entropy), periodic in the space
variables. In [An1] it is next proved that 
the local J\"{u}ttner solution must be a global one, due to the
periodicity of the solution. 

For more information on the relativistic Boltzmann equation on 
Minkowski space we refer to [Gl], [GLW] and [Sy] and in the nonrelativistic
case we refer to the excellent review paper by Villani [V] and the
books [Gl], [Ci]. 
\subsection{The Vlasov-Maxwell- and Vlasov-Poisson systems} 
Let us consider a collisionless plasma which is a collection of 
particles where collisions
are relatively rare and the interaction is through 
their charges. We assume below that the plasma only consist of one type of
particles whereas the results below also hold for plasmas with several
particle species. 
The particle rest mass is normalized to one. 
In the kinetic framework, the most general set of equations for modelling a
collisionless plasma is the relativistic
Vlasov-Maxwell system 
\begin{equation}
\partial_{t}f+\hat{v}\cdot\nabla_{x}f+(E(t,x)+\hat{v}\times
B(t,x))\cdot\nabla_{v}f=0\label{rvm}
\end{equation}
\begin{eqnarray}
  \label{maxwell}
  &\partial_{t}E+j=c\nabla\times B,\;\;\;\;\;\nabla\cdot E=\rho, &\\ 
  &\partial_{t}B=-c\nabla\times E,\;\;\;\;\;\nabla\cdot B=0.&\label{maxwell_2}
\end{eqnarray}
The notation follows the one already introduced with the exception
that the momenta are now denoted by $v$ instead of $p$. This has become
a standard notation in this field. 
$E$ and $B$ are the
electric and magnetic fields, and $\hat{v}$ is the
relativistic velocity 
\begin{equation}
\hat{v}=\frac{v}{\sqrt{1+|v|^2/c^2}},
\label{relveldef}
\end{equation}
where $c$ is the speed
of light. 
The charge density $\rho$ and current $j$ are given by 
\begin{equation}
  \rho=\int_{\mathbb{R}^{3}} fdv,\;\;\;\;\; j=\int_{\mathbb{R}^{3}}
\hat{v}fdv. 
\end{equation}
Equation (\ref{rvm}) is the relativistic Vlasov equation and 
(\ref{maxwell}) and (\ref{maxwell_2}) are the Maxwell equations. 

A special case in three dimensions is obtained by considering 
spherically symmetric initial data. For such data it can be shown 
that also the solution will be spherically symmetric, and that 
the magnetic field has to be constant. The Maxwell equation 
$\nabla\times E=-\partial_tB$ then implies that the electric field 
is the gradient of a potential $\phi$. 
Hence, in the spherically
symmetric case the relativistic Vlasov-Maxwell system 
takes the form, 
\begin{eqnarray}
&\partial_{t}f+\hat{v}\cdot\nabla_{x}f+\beta E(t,x)
\cdot\nabla_{v}f=0&\label{rv}\\
&E=\nabla\phi,\;\;\;\Delta\phi=\rho.&\label{rpoisson}
\end{eqnarray}
Here $\beta=1,$ and the constant magnetic field has been 
set to zero, since a constant field has no significance in 
this discussion. 
This system makes sense
for any initial data, without symmetry constraints, and 
is called the relativistic Vlasov-Poisson equation. 
Another 
special case of interest is the classical limit, 
obtained by letting $c\rightarrow\infty$ in
(\ref{rvm})-(\ref{maxwell_2}), 
\begin{eqnarray}
&\partial_{t}f+v\cdot\nabla_{x}f+\beta E(t,x)
\cdot\nabla_{v}f=0&\label{v}\\
&E=\nabla\phi,\;\;\;\Delta\phi=\rho,&\label{poisson}
\end{eqnarray}
where $\beta=1.$ 
See Schaeffer [Sc1] for a rigorous derivation of this result. 
This is the (nonrelativistic) Vlasov-Poisson equation, 
and $\beta=1$ corresponds to repulsive 
forces (the plasma case). Taking $\beta=-1$ means attractive forces
and the Vlasov-Poisson equation is then a model for a Newtonian
self-gravitating system. 

One of the fundamental problems in kinetic theory is to find out
whether or not spontaneous shock formations will develop in a
collisionless gas, i.e. if solutions 
to any of the equations above will remain smooth 
for all 
time, given smooth initial data. 

If the initial data are small this problem has an affirmative solution
in all 
cases considered above (see [GSc1], [GSt2], [BD], and [BDH]). 
For initial data unrestricted in size the picture is more involved. 
In order to 
obtain smooth solutions globally in time, the main issue is to 
control the support of the momenta 
\begin{equation}
Q(t):=\sup \{|v|:\exists (s,x)\in [0,t]\times \mathbb{R}^3\mbox{ such that
    }f(s,x,v)\not= 0\},\label{Q}
\end{equation} 
i.e. to bound $Q(t)$ by a continuous function so that $Q(t)$ will not
blow up in finite time. 
That such a control is sufficient for obtaining global existence of
smooth solutions follows from well-known results in the different
cases, see [GSt1], [Ho1], [B] and [GSc1]. 
For the full three-dimensional relativistic Vlasov-Maxwell system this
important problem of establishing whether or not solutions will remain
smooth for all time is open. In two space- and three momentum
dimensions Glassey and Schaeffer [GSc2] have shown that $Q(t)$ can be
controlled which thus yields global existence of smooth solutions 
in that case (see also [GSc3]). 

The relativistic and nonrelativistic Vlasov-Poisson equation 
are very similar in form. 
In particular, the equation for the field is 
identical in the two cases. 
However, the mathematical results about the two systems are 
very different. 
In the nonrelativistic case Batt [B] gave an affirmative solution 1977
in the case of spherically symmetric data. Pfaffelmoser [Pf] was the
first one to give a proof for general smooth data. He obtained the bound 
$$Q(t)\leq C(1+t)^{(51+\delta)/11},$$ where $\delta>0$ could be taken
arbitrary small. This bound was later improved by different authors. 
The sharpest bound valid 
for $\beta=1$ and $\beta=-1$ has been given by Horst [Ho2] and reads 
$$Q(t)\leq C(1+t)\log(2+t).$$ In the case of repulsive forces 
($\beta=1$) Rein [Rn1] has found the sharpest estimate by using a new
identity for the Vlasov-Poisson equation, discovered independently 
by Illner and Rein [IlR] and by Perthame [Pe]. Rein's estimate reads 
$$Q(t)\leq C(1+t)^{2/3}.$$ Independently and about the same time as 
Pfaffelmoser gave his proof, Lions and Perthame [LP] used a different
method for proving global existence. To some extent their method
seems to be more generally applicable to attack problems similar to
the Vlasov-Poisson equation but which are still quite different 
(see [An2], [KR]). On the other hand their method does not give such
strong growth estimates on $Q(t)$ as described above. 
For the relativistic Vlasov-Poisson equation (with $\beta=1$) 
Glassey and Schaeffer [GSc1] showed that if the data are spherically 
symmetric $Q(t)$ can be controlled, which is analogous to 
the result by Batt mentioned above (we also mention that the
cylindrical case has been considered in [GSc4]). If $\beta=-1$ it was
also shown in [GSc1] 
that blow up occurs in finite time for spherically symmetric
data with negative total energy. This system is however unphysical in the sense that it is not a
special case of the Einstein-Vlasov system. 
Quite surprisingly, for general smooth initial data 
none of the techniques discussed above for the
nonrelativistic Vlasov-Poisson equation apply in the relativistic case. 
This fact is annoying since it has 
been suggested 
that an understanding of this equation may be necessary 
for understanding the three-dimensional relativistic Vlasov-Maxwell 
equation. However, the relativistic Vlasov-Poisson equation 
lacks the Lorentz invariance,
it is a hybrid of a classical Galilei invariant field equation and a
relativistic transport equation (\ref{rv}). Only for spherical
symmetric data the equation is a fundamental 
physical equation. The classical Vlasov-Poisson equation 
is on the other hand Galilean invariant. 
In 
[An3] a different equation for the field is introduced which is observer 
independent
among Lorentz observers. 
By coupling this equation for the field 
to the relativistic Vlasov equation the function $Q(t)$ 
may be controlled as shown in [An3]. 
This is an indication that the transformation
properties are important in studying existence of smooth
solutions (the situation is less subtle for weak solutions, 
where 
energy estimates and averaging are the main tools, see [HH] and
[DL2]). 
Hence, it
is unclear whether or not the relativistic Vlasov-Poisson equation will 
play a central role 
in the understanding of the Lorentz invariant relativistic 
Vlasov-Maxwell equation. 

We refer to the book by Glassey [Gl] for more information on the
relativistic Vlasov-Maxwell system and the Vlasov-Poisson equation. 
\subsection{The Einstein-Vlasov system} 
In this section we will consider a self-gravitating collisionless gas
and we will write down the Einstein-Vlasov system and describe its
general mathematical features. Our 
presentation follows to a large extent the one by Rendall in [Rl6]. We 
also refer to 
Ehlers [E] and [St] for more background on kinetic theory in general
relativity. 

Let $M$ be a four-dimensional manifold and let $g_{ab}$ be a metric
with Lorentz signature $(-,+,+,+)$ so that $(M,g_{ab})$ is a
spacetime. We use the abstract index notation which means that
$g_{ab}$ is a geometric object and not the components of a tensor. See
[Wa] for a discussion on this notation. The metric is assumed to be
time-orientable so that there is a distinction between future and past
directed vectors. The worldline of a particle with non-zero rest mass $m$ 
is a timelike curve and the unit future directed tangent vector $v^a$
to this curve is the four-velocity of the particle. The four-momentum $p^a$ is
given by $mv^a.$  We assume that all particles have equal rest mass
$m$ and we normalize so that $m=1$. One can also consider massless
particles but we will rarely discuss this case. The possible values of
the four-momentum are all future directed unit timelike vectors and
they constitute a hypersurface $P$ in the tangent bundle $TM$ which is 
called the mass shell. The distribution function $f$ that we introduced in
the previous sections is a non-negative function on $P.$ Since we are
considering a collisionless gas the particles travel along geodesics
in spacetime. The Vlasov equation is an equation for $f$ which exactly 
expresses this fact. To get an explicit expression for this equation
we introduce local coordinates on the mass shell. We choose local
coordinates on $M$ such that the hypersurfaces $t=x^0=$constant are 
spacelike so that $t$ is a time coordinate and $x^j,\;j=1,2,3$ are
spatial coordinates (letters in the beginning of the alphabet always
take values $0,1,2,3$ and letters in the middle take $1,2,3$). 
A timelike vector is future directed if and only if its zero component
is positive. Local coordinates on $P$ can then 
be taken as $x^a$ together with the spatial components of the 
four-momentum $p^a$ in these coordinates. 
The Vlasov equation then reads 
\begin{equation}
\partial_{t}f+\frac{p^{j}}{p^0}\partial_{x^j}f
-\frac{1}{p^0}\Gamma^{j}_{ab}p^ap^b\partial_{p^j}f=0.\label{vlasovgamma} 
\end{equation} 
Here $a,b=0,1,2,3$ and $j=1,2,3$ and $\Gamma^{j}_{ab}$ are the
Christoffel symbols. It is understood that $p^0$ is expressed in terms
of $p^j$ and the metric $g_{ab}$ using the relation $g_{ab}p^ap^b=-1$
(recall that $m=1$). 

In a fixed spacetime the Vlasov equation is a linear hyperbolic
equation for $f$ and we can solve it by solving the characteristic
system 
\begin{eqnarray} 
dX^i/ds&=&\frac{P^i}{P^0},\label{char1}\\ 
dP^i/ds&=&-\Gamma^i_{ab}\frac{P^aP^b}{P^0}.\label{char2} 
\end{eqnarray} 
In terms of initial data $f_0$ the solution to the Vlasov equation can
be written 
\begin{equation}
f(x^a,p^i)=f_0(X^i(0,x^a,p^i),P^i(0,x^a,p^i)),\label{solution} 
\end{equation} 
where $X^i(s,x^a,p^i)$ and $P^i(s,x^a,p^i)$ solve (\ref{char1}) and
(\ref{char2}) and where $X^i(t,x^a,p^i)=x^i$ and $P^i(t,x^a,p^i)=p^i.$ 

In order to write down the Einstein-Vlasov system we need to define
the energy-momentum tensor $T_{ab}$ in terms of $f$ and $g_{ab}.$ 
In the coordinates $(x^a,p^a)$ on $P$ we define 
\begin{displaymath} 
T_{ab}=-\int_{\mathbb{R}^{3}}f\,p_ap_b|g|^{1/2}\frac{dp^1dp^2dp^3}{p_0}, 
\end{displaymath} 
where as usual $p_a=g_{ab}p^b$ and $|g|$ denotes the absolute value of
the determinant of $g.$ Equation (\ref{vlasovgamma}) together with 
Einstein's equations 
\begin{displaymath} 
G_{ab}:=R_{ab}-\frac{1}{2}Rg_{ab}=8\pi T_{ab}, 
\end{displaymath} 
then form the Einstein-Vlasov system. 
Here $G_{ab}$ is the Einstein tensor, $R_{ab}$ the Ricci tensor and $R$
is the scalar curvature. 
We also define the particle current density 
\begin{displaymath} 
N^a=-\int_{\mathbb{R}^{3}}f\,p^a|g|^{1/2}\frac{dp^1dp^2dp^3}{p_0}. 
\end{displaymath} 
Using normal coordinates based at a given point and assuming that $f$
is compactly supported it is not hard to see
that $T_{ab}$ is divergence-free which is a necessary compatability
condition since $G_{ab}$ is divergence-free by the Bianchi identities. 
A computation in normal coordinates also shows that $N^a$ is
divergence-free which expresses the fact that the number of particles
is conserved. The definitions of $T_{ab}$ and $N^a$ immediately give us
a number of inequalities. If $V^a$ is a future directed timelike or
null vector then we have $N_aV^a\leq 0$ with equality if and only if
$f=0$ at the given point. Hence $N^a$ is always future directed
timelike if there are particles at that point. 
Moreover, if $V^a$ and $W^a$ are future directed timelike vectors then
$T_{ab}V^aW^b\geq 0,$ which is the dominant energy condition. 
If $X^a$ is a spacelike vector then $T_{ab}X^aX^b\geq 0.$ This is 
called the non-negative pressure condition. These last two conditions
together with the Einstein equations imply that $R_{ab}V^aV^b\geq 0$
for any timelike vector $V^a,$ which is the strong energy condition. 
That the energy conditions hold for Vlasov matter is one reason that
the Vlasov equation defines a well-behaved matter model in general
relativity. Another reason is the well-posedness theorem by
Choquet-Bruhat for the Einstein-Vlasov system that we will state below. 
Before stating that theorem we will first 
discuss the initial conditions imposed. 

The data in the Cauchy problem for the Einstein-Vlasov system consist
of the induced Riemannian metric $g_{ij}$ on the initial hypersurface $S$, the
second fundamental form $k_{ij}$ of $S$ and matter data $f_0$. 
The relations between a given initial data set 
$(g_{ij},k_{ij})$ on a three-dimensional manifold $S$ and the metric $g_{ij}$ 
on the spacetime manifold is that there exists an imbedding $\psi$ of $S$ 
into the spacetime such that the induced metric and second fundamental form of 
$\psi(S)$ coincide with the result of transporting $(g_{ij},k_{ij})$
with $\psi$. 
For the relation of the distribution functions $f$ and $f_0$ we have to 
note that $f$ is defined on the mass shell. The initial condition
imposed is that the restriction of $f$ to the part of the mass shell
over $\psi(S)$  
should be equal to $f_0\circ (\psi^{-1},d(\psi)^{-1})\circ\phi$ where $\phi$ 
sends each point of the mass shell over $\psi(S)$, to its orthogonal 
projection onto the tangent space to $\psi(S)$. An initial data 
set for the Einstein-Vlasov system must satisfy the constraint 
equations which read 
\begin{eqnarray}
R-k_{ij}k^{ij}+(trk)^2&=&16\pi\rho,\label{constr1}\\ 
\nabla_{i}k^{i}_{l}-\nabla_{l}(trk)&=&8\pi j_l.\label{constr2} 
\end{eqnarray} 
Here $\rho=T_{ab}n^an^b$ and $j^a=-h^{ab}T_{bc}n^c$ where $n^a$ is the
future directed unit normal vector to the initial hypersurface and
$h^{ab}=g^{ab}+n^an^b$ is the orthogonal projection onto the tangent
space to the initial hypersurface. In terms of $f_0$ we can express
$\rho$ and $j^l$ by ($j^a$ satisfies $n_aj^a=0$ so it can naturally be
identified with a vector intrinsic to $S$) 
\begin{eqnarray*}
\rho&=&\int_{\mathbb{R}^{3}}f\,p^ap_a|^{(3)}g|^{1/2}\frac{dp^1dp^2dp^3}
{1+p_jp^j},\\ 
j_l&=&\int_{\mathbb{R}^{3}}f\,p_l|^{(3)}g|^{1/2}\, dp^1dp^2dp^3. 
\end{eqnarray*} 
Here $|^{(3)}g|$ is the determinant of the induced Riemannian metric
on $S.$ 
We can now state the local existence theorem by Choquet-Bruhat [Ct] 
for the Einstein-Vlasov system. 
\begin{theorem} 
Let $S$ be a 3-dimensional manifold, $g_{ij}$ a smooth Riemannian
metric on $S,$ $k_{ij}$ a smooth symmetric tensor on $S$ and $f_0$ a
smooth non-negative function of compact support on the tangent bundle
$TS$ of $S$. Suppose that these objects satisfy the constraint
equations (\ref{constr1}) and (\ref{constr2}). Then there exists a
smooth spacetime $(M,g_{ab}),$ a smooth distribution function $f$ on
the mass shell of this spacetime and a smooth embedding $\psi$ of $S$
into $M$ which induces the given initial data on $S$ such that
$g_{ab}$ and $f$ satisfy the Einstein-Vlasov system and $\psi(S)$ is a
Cauchy surface. Moreover, given any other spacetime $(M',g'_{ab}),$
distribution function $f'$ and embedding $\psi'$ satisfying these
conditions, there exists a diffeomorphism $\chi$ from an open
neighbourhood of $\psi(S)$ in $M$ to an open neighbourhood of
$\psi'(S)$ in $M'$ which satisfies $\chi\circ\psi=\psi'$ and carries
$g_{ab}$ and $f$ to $g'_{ab}$ and $f'$ respectively.  
\end{theorem} 
In this context we also mention that local existence has been proved
for the Einstein-Maxwell-Boltzmann system [BC] and for the
Yang-Mills-Vlasov system [CN]. 

A main theme in the following sections is to discuss special cases
when the local existence theorem can be extended to a global
one. There are interesting situations when this can be achieved and
such global existence theorems are not known for Einstein's equations
coupled to other forms of phenomenological matter models, e.g. fluid
models (see however [Cu3]). 
In this context it should be stressed that the results in the
previous sections show that the mathematical understanding of kinetic
equations on a flat background space is well-developed. On the other
hand the mathematical understanding of fluid
equations on a flat background space (also in the absence of a
Newtonian gravitational field) is not that well-understood. It 
would be desirable to have a better mathematical understanding of
these equations in the absence of gravity before coupling them to
Einstein's equations. This suggests that the Vlasov equation is
natural as matter model in mathematical general relativity. 
\section{Global existence theorems for the Einstein-Vlasov system} 
In general relativity two classes of initial data are distinguished. 
If an isolated body is studied the data are called asymptotically
flat. The initial hypersurface is topologically $\mathbb{R}^{3}$ and far
away from the body one expects spacetime to be approximately 
flat and appropriate fall off conditions are
imposed. Roughly, a 
smooth data set $(g_{ij},k_{ij},f_0)$ on $\mathbb{R}^{3}$ is said to be
asymptotically flat 
if there exist global coordinates $x^i$ such that as $|x|$ tends to
infinity the components $g_{ij}$ in these coordinates tend to
$\delta_{ij},$ the components $k_{ij}$ tend to zero, $f_0$ has
compact support and certain weighted Sobolev norms of
$g_{ij}-\delta_{ij}$ and $k_{ij}$ are finite (see [Rl6]). 
The symmetry classes that admit asymptotically flatness are few. 
The important ones are spherically symmetric and axially symmetric 
spacetimes. One can also consider a case where spacetime is
asymptotically flat except in one direction, namely cylindrical
spacetimes. Regarding global existence questions only spherically 
symmetric spacetimes have been considered for the Einstein-Vlasov
system in the asymptotically flat case. 

Spacetimes that possess a compact Cauchy hypersurface are called
cosmological spacetimes and data are accordingly given on a compact
3-manifold. In this case the whole universe is modelled and not only
an isolated body. In contrast to the asymptotically flat case
cosmological spacetimes admit a large number of symmetry classes. This 
gives a possibility to study interesting special cases for which the 
difficulties of the full Einstein equations are strongly reduced. 
We will discuss below cases where the spacetime is characterized by 
the dimension of its isometry group together with the dimension of 
the orbit of the isometry group. 

\subsection{Spherically symmetric spacetimes} 
The study of the global properties of solutions to the spherically
symmetric Einstein-Vlasov system was initiated by Rein and Rendall in
1990. 
They choose to work in coordinates where the metric takes the form 
\begin{displaymath}
ds^{2}=-e^{2\mu(t,r)}dt^{2}+e^{2\lambda(t,r)}dr^{2}
+r^{2}(d\theta^{2}+\sin^{2}{\theta}d\varphi^{2}), 
\end{displaymath} 
where $t\in\mathbb{R},\, r\geq 0,\,\theta\in [0,\pi],\,\varphi\in
[0,2\pi].$ These are called Schwarzschild coordinates. 
Asymptotic flatness is expressed by the boundary conditions 
\begin{displaymath} 
\lim_{r\rightarrow\infty}\lambda(t,r)=\lim_{r\rightarrow\infty}\mu(t,r)
=0,\;\forall t\geq 0. 
\end{displaymath} 
A regular centre is also required and is guaranteed by the boundary
condition 
\begin{displaymath} 
\lambda(t,0)=0. 
\end{displaymath} 
With $$x=(r\sin\phi\cos\theta,r\sin\phi\sin\theta,r\cos\phi)$$ as 
spatial coordinate and $$v^j=p^j+(e^\lambda-1)\frac{x\cdot
  p}{r}\frac{x^j}{r}$$ as momentum coordinates the Einstein-Vlasov
system reads 
\begin{equation} 
\partial_{t}f+e^{\mu-\lambda}\frac{v}{\sqrt{1+v^2}}\cdot\partial_{x}f
-(\lambda_{t}\frac{x\cdot v}{r}+e^{\mu-\lambda}\mu_{r}\sqrt{1+v^2})
\frac{x}{r}\cdot\partial_{v}f=0,\label{Vlas} 
\end{equation} 
\begin{eqnarray} 
&\displaystyle e^{-2\lambda}(2r\lambda_{r}-1)+1=8\pi
r^2\rho,&\label{ee1}\\ 
&\displaystyle e^{-2\lambda}(2r\mu_{r}+1)-1=8\pi r^2 p.&\label{ee2} 
\end{eqnarray} 
The matter quantities are defined by 
\begin{eqnarray} 
\rho(t,x)&=&
\int_{\mathbb{R}^{3}}\sqrt{1+|v|^2}f(t,x,v)\;dv,\label{rho}\\ 
p(t,x)&=&\int_{\mathbb{R}^{3}}\left(\frac{x\cdot
    v}{r}\right)^{2}f(t,x,v)\;\frac{dv}{\sqrt{1+|v|^2}}.\label{p} 
\end{eqnarray} 
Let us point out that this system is not the full Einstein-Vlasov
system. The remaining field equations can however be derived from
these equations. See [RR1] for more details. 
Let the square of the angular momentum be denoted by $L,$ i.e. 
$$L:=|x|^2|v|^2-(x\cdot v)^2.$$ 
A consequence of spherical symmetry is that angular momentum 
is conserved along the characteristics of (\ref{Vlas}). 
Introducing the variable 
$$w=\frac{x\cdot v}{r},$$ the Vlasov equation for $f=f(t,r,w,L)$ 
becomes 
\begin{equation} 
\partial_{t}f+e^{\mu-\lambda}\frac{w}{E}\partial_{r}f
-(\lambda_{t}w+e^{\mu-\lambda}\mu_{r}E-
e^{\mu-\lambda}\frac{F}{r^3E})\partial_{w}f=0,\label{Vlas2} 
\end{equation} 
where 
$$E=E(r,w,L)=\sqrt{1+w^{2}+L/r^{2}}.$$ 
The matter terms take the form 
\begin{eqnarray} 
\rho(t,r)&=&\frac{\pi}{r^{2}}
\int_{-\infty}^{\infty}\int_{0}^{\infty}Ef(t,r,w,L)\;dwdL,\label{rho2}\\ 
p(t,r)&=&\frac{\pi}{r^{2}}\int_{-\infty}^{\infty}\int_{0}^{\infty}
\frac{w^{2}}{E}f(t,r,w,L)\;d
wdL.\label{p2} 
%j(t,r)&=&\frac{\pi}{r^{2}}
%\int_{-\infty}^{\infty}\int_{0}^{\infty}wf(t,r,w,F),\;dwdF\label{j}\\ 
%q(t,r)&=&\frac{\pi}{r^{2}}\int_{-\infty}^{\infty}\int_{0}^{\infty}\frac{F}{E}f(t,r,w,F)\;
%dwdF. 
\end{eqnarray} 
Let us write down a couple of known 
facts about the system
(\ref{ee1}),(\ref{ee2}),(\ref{Vlas2}),(\ref{rho2}) and (\ref{p2}). 
A solution to the Vlasov 
equation can be written 
\begin{equation}
f(t,r,w,L)=f_{0}(R(0,t,r,w,L),W(0,t,r,w,L),L), 
\label{repre} 
\end{equation} 
where $R$ and $W$ are solutions to the characteristic system
\begin{eqnarray}
\frac{dR}{ds}&=&e^{(\mu-\lambda)(s,R)}\frac{W}{E(R,W,L)},\label{char1}\\ 
\frac{dW}{ds}&=&-\lambda_{t}(s,R)W-e^{(\mu-\lambda)(s,R)}\mu_{r}(s,R)E(R,W,L)
\label{char2}\\ 
&\phantom{hej}&+e^{(\mu-\lambda)(s,R)}\frac{L}{R^3E(R,W,L)},\nonumber 
\end{eqnarray} 
such that the trajectory $(R(s,t,r,w,L),W(s,t,r,w,L),L)$ goes 
through the point $(r,w,L)$ when $s=t$. 
This representation shows that $f$ is nonnegative for all $t\geq 0$ and that 
$f\leq\|f_0\|_{\infty}.$ 
There are two known conservation laws for the Einstein-Vlasov
system, conservation of the number of particles 
\begin{displaymath} 
4\pi^{2}\int_{0}^{\infty}e^{\lambda(t,r)}
\left(\int_{-\infty}^{\infty}\int_{0}^{\infty}f(t,r,w,L)dLdw\right) dr, 
\end{displaymath} 
and conservation of the ADM mass 
\begin{equation} 
M:=4\pi\int_{0}^{\infty}r^{2}\rho(t,r)dr.\label{adm} 
\end{equation} 

Let us now review the global results concerning the Cauchy problem
that have been proved for 
the spherically symmetric Einstein-Vlasov system. As initial data we
take a spherically symmetric, nonnegative and 
continuously differentiable function $f_0$ with compact support which
satisfies 
\begin{displaymath} 
4\pi^{2}\int_{0}^{r}\int_{-\infty}^{\infty}\int_{0}^{\infty}
Ef_{0}(r,w,L)dwdLdr<\frac{r}{2}. 
\end{displaymath} 
This condition guarantees that no trapped surfaces are present initially. 
In [RR1] it is shown that for such an initial datum there exists a
unique, continuously differentiable solution $f$ with $f(0)=f_0$ on
some right maximal interval $[0,T).$ If the solution blows up in
finite time, i.e. if $T<\infty$ then $\rho(t)$ becomes unbounded as
$t\rightarrow T.$ Moreover, a continuation criterion is shown which
says that a local solution can be extended to a global one provided 
the $v$-support of $f$ can be bounded on $[0,T)$ (in [RR1] they 
choose to work in the momentum variable $v$ rather than $w,L.$). 
This is analogous to the situation for the Vlasov-Maxwell system where
the function $Q(t)$ was introduced for the $v$-support. A control of
the $v$-support immediately implies that $\rho$ and $p$ are bounded in
view of (\ref{rho}) and (\ref{p}). In the Vlasov-Maxwell case the
field equations have a regularizing effect in the sense that 
derivatives can be expressed through spatial integrals, and it follows 
[GSt1] that also the derivatives of $f$ can be bounded if the 
$v$-support is bounded. For 
the Einstein-Vlasov system such a regularization is less clear, 
e.g. $\mu_r$ depends on $\rho$ in a pointwise manner. However, certain 
combinations of second and first order derivatives of the metric
components can be expressed in
terms of matter components only and no derivatives of those (a
consequence of the geodesic deviation equation). This 
fact turns out to be sufficient for obtaining bounds
also on the derivatives of $f$ (see [RR1] for details). 
By considering initial data sufficiently close to zero, Rein 
and Rendall show that the $v$-support is bounded on 
$[0,T)$ and the continuation criterion then implies that $T=\infty.$ It
should be stressed that even for small data no global 
existence result like this one is known for any other phenomenological
matter model coupled to Einstein's equations. 
The resulting spacetime in [RR1] is geodesically 
complete and the components of the energy momentum tensor
as well as the metric quantities decay with certain algebraic rates in
$t.$ The mathematical method used by Rein and Rendall is inspired 
by the analogous small data result for the Vlasov-Poisson equation by 
Bardos and Degond [BD]. This should not be too surprising since for
small data the gravitational fields are expected to be small and a
Newtonian spacetime should be a fair approximation. In this context we
point out that in [RR2] it is proved that the Vlasov-Poisson
system is indeed the nonrelativistic limit of the spherically
symmetric Einstein-Vlasov system,
i.e. the limit when the speed of light $c\rightarrow \infty.$ 
(In [Rl3] this issue is studied in the asymptotically flat case
without symmetry assumptions.) 
Finally we mention that there is an analogous small data result using
a maximal time coordinate [Rl6] instead of a Schwarzschild time coordinate. 

The case with general data is more subtle. 
Rendall has shown [Rl7] that there exists data leading to singular
spacetimes as a consequence of Penrose's singularity theorem. 
This raises the question what we mean by global existence for such 
data. The Schwarzschild time coordinate is expected to avoid the
singularity and by global existence we mean that solutions remain
smooth as Schwarzschild time tends to infinity. Even though spacetime 
might only be partially covered in Schwarzschild coordinates a global
existence theorem for general data would nevertheless 
be very important since weak cosmic censorship would follow from it. 
A partial investigation for general data was done in 
[RRS1] where it is shown that 
if singularities form in finite Schwarzschild time the first one must
be at the centre. More precisely, if $f(t,r,w,L)=0$ when $r>\epsilon$
for some 
$\epsilon>0$ for all $t,w$ and $L,$ then the solution remains smooth for all
time. This rules out singularities of shell crossing type which can
be an annoying problem for other matter models (e.g. dust). The main
observation 
in [RRS1] is a cancellation property in the term $$\lambda_t w+
e^{\mu-\lambda}\mu_rE$$ in the characteristic equation (\ref{char2}). 
We refer to the original paper for details. 
In [RRS2] a numerical study was undertaken. 
A numerical scheme originally used for the Vlasov-Poisson system 
was modified to the spherically symmetric Einstein-Vlasov system. 
It has been shown by Rodewis [Rs] that the numerical scheme has 
the desirable convergence properties. (In the Vlasov-Poisson case
convergence was proved in [Sc3], see also [Ga]). 
The numerical experiments support the conjecture 
that solutions are 
singularity-free. This can be seen as evidence that weak cosmic
censorship holds for collisionless matter. It may even hold in
a stronger sense than in the case of a massless scalar field (see
[Cu1-2]). There may be no naked singularities formed for any regular
initial data rather than just for generic data. This speculation is
based on the fact that 
the naked singularities which occur in scalar field collapse appear to
be associated with the existence of type II critical collapse while
Vlasov matter is of type I. This is indeed the primary goal of 
their numerical investigation to analyze critical collapse and
decide whether Vlasov matter is type I or type II. These 
different types of matter are defined as follows. Given small initial
data no black holes are expected to form and matter will disperse
(which have been proved for a scalar field [Cu4] and for Vlasov
matter [RR1]). 
For large data black holes will form and consequently there
is a transition regime separating 
dispersion of matter and formation of black holes. If we introduce a 
parameter $A$ on the initial data such that for small $A$ dispersion
occurs and for large $A$ a black hole is formed we get a critical value
$A_c$ separating these regions. If we take $A>A_c$ and denote by
$m_B(A)$ the mass of the black hole then if 
$m_B(A)\rightarrow 0$ as $A\rightarrow A_c,$ we have type II matter
whereas for type I matter 
this limit is positive and there is a mass gap. For more 
information on critical collapse we refer to the review paper by 
Gundlach [Gu]. For Vlasov matter there is an independent numerical
simulation by Olabarrieta and Choptuik [OC] (using a maximal time
coordinate) and their conclusion agrees with the one in [RRS2]. 
Critical collapse is related to self 
similar solutions and Martin-Garcia and Gundlach [MG] have presented a
construction of such solutions for the massless Einstein-Vlasov system 
by using a method based partially on numerics. Since such solutions
often are related to naked singularities it is important to note that
their result is for the massless case (in which case there is no known
analogous result to the small data theorem in [RR1]) and their initial
data are not in the class that we have described above. 
\subsection{Cosmological solutions} 
In cosmology the whole universe is modelled and the ``particles'' in
the kinetic description are galaxies or even clusters of galaxies. 
The main goal is again to determine the global properties of the solutions
to the Einstein-Vlasov system. In order to do so a 
global time coordinate $t$ must be found (global existence) and 
the asymptotic behaviour of the solutions when $t$ tends to its
limiting values has to be analyzed. This might correspond to 
approaching a singularity (eg. the big bang singularity) or to a phase
of unending expansion. Since the general case is beyond the range of
current mathematical techniques all known results are for cases 
with symmetry (see however [Ae] where to some extent global properties 
are established in the case without symmetry). 

There are several existing results on global time coordinates for 
solutions of the Einstein-Vlasov system. In the spatially homogeneous 
case it is natural to choose a Gaussian time coordinate based on
a homogeneous hypersurface. The maximal range of a Gaussian time
coordinate in a solution of the Einstein-Vlasov 
system evolving from data on a compact manifold which are homogeneous 
was determined in [Rl4]. It is 
finite for models of Bianchi IX and Kantowski-Sachs types and finite in 
one time direction and infinite in the other for the other Bianchi types. 
The asymptotic behaviour of solutions in the spatially homogeneous
case has been analyzed in [RT] and [RU]. In [RT] the case of massless
particles is considered whereas the massive case is studied in [RU]. 
Both the nature of the initial singularity and the phase of unlimited
expansion is analyzed. The main concern is about
Bianchi models I,II and III. The authors compare their solutions
with the solutions to the corresponding perfect fluid models. A
general conclusion is that the choice of
matter model is very important since for all symmetry classes studied 
there are differences between the collisionless model and a perfect fluid
model both regarding the initial singularity and the expanding phase. 
The most striking example is for the Bianchi II models where they find
persistent oscillatory behaviour near the singularity which is quite
different from the known behaviour of Bianchi type II perfect fluid
models. 
In [RU] it is also shown that solutions for massive particles are
asymptotic to solutions with massless particles near the initial
singularity. For Bianchi I and II it is also proved that solutions
with massive particles are asymptotic to dust solutions at late times.
It is conjectured that the same holds true also for Bianchi III. This
problem is then settled by Rendall in [Rl8]. 

All other results presently available on the subject concern spacetimes 
which admit a group of isometries acting on two-dimensional spacelike 
orbits, at least after passing to a covering manifold. 
The group may be 
two-dimensional (local $U(1)\times U(1)$ or $T^2$ symmetry) 
or three-dimensional 
(spherical, plane or hyperbolic symmetry). In all these cases the 
quotient of spacetime by the symmetry group has the structure of a 
two-dimensional Lorentzian manifold $Q$. The orbits of the group action 
(or appropriate quotients in the case of a local symmetry) are called 
surfaces of symmetry. Thus there is a one-to-one correspondence between 
surfaces of symmetry and points of $Q$.
There is a major difference between the cases where the symmetry group
is two- or three-dimensional. In the three-dimensional case 
no gravitational waves are admitted in contrast to the two-dimensional
case. In the former case the field equations reduce to ODEs while in
the latter their evolution part consists of nonlinear wave equations. 
Three types of time coordinates which have been studied
in the inhomogeneous case are CMC, areal and conformal 
coordinates. A CMC time coordinate $t$ is one where each hypersurface 
of constant time has constant mean curvature (CMC) and on each hypersurface 
of this kind the value of $t$ is the mean curvature of that slice.
In the case of areal coordinates the time coordinate is a function of
the area of the surfaces of symmetry (eg. proportional to the area or 
proportional to the square root of the area). In the case of conformal  
coordinates the metric on the quotient manifold $Q$ is conformally
flat. 

Let us first consider spacetimes $(M,g)$ admitting a three-dimensional
group of isometries. 
The topology of $M$ is assumed to be 
$\mathbb{R}\times S^1\times F,$ with $F$ a compact two-dimensional 
manifold. The universal covering $\hat{F}$ of $F$ induces a spacetime 
$(\hat{M},\hat{g})$ by $\hat{M}=\mathbb{R}\times S^1\times\hat{F}$ and
$\hat{g}=p^{*}g$ where $p:\hat{M}\rightarrow M$ is the canonical
projection. A three-dimensional group $G$ of isometries is assumed to
act on $(\hat{M},\hat{g}).$ If $F=S^2$ and $G=SO(3)$ then $(M,g)$ is
called spherically symmetric, if $F=T^2$ and $G=E_2$ (Euclidean group)
then $(M,g)$ is called plane symmetric, and if $F$ has genus greater 
than one and the connected component of the symmetry group $G$ of the
hyperbolic plane $H^2$ acts isometrically on $\hat{F}=H^2$ then
$(M,g)$ is said to have hyperbolic symmetry. 

In the case of spherical symmetry the existence of one compact CMC 
hypersurface implies that the whole spacetime can be covered by a 
CMC time coordinate which takes all real values [Rl1,BR]. 
The existence of one compact CMC hypersurface in this case
was proved later by Henkel [He1] using the concept of 
prescribed mean curvature (PMC) foliation. 
Accordingly this gives a complete picture in the spherically 
symmetric case regarding CMC foliations. 
In the case of areal coordinates Rein [Rn2] has shown, under a size
restriction on the initial data, 
that the past of an initial hypersurface can be covered. In
the future direction it is shown that areal coordinates break down in
finite time. 

In the case of plane and hyperbolic symmetry Rendall and 
Rein showed in [Rl1] and [Rn2] respectively, 
that the existence results (for CMC time
and areal time) 
in the past direction for spherical symmetry also hold for 
these symmetry classes. The global CMC foliation results 
to the past imply that the singularity is a crushing singularity since
the mean curvature blows up at the singularity. In addition, Rein also
proved in his special case with small initial data that 
the Kretschmann curvature scalar blows up when the singularity is
approached. 
Hence the singularity is both a crushing and a curvature singularity
in this case. 
In both of these works the question of global existence to 
the future was left open.
This gap was closed by the author, Rein and Rendall in [ARR] and global 
existence to the future was established in both CMC- and areal 
coordinates. The global existence result for CMC time is partly a 
consequence of the global existence theorem in areal coordinates
together with a theorem by Henkel [He1] which shows that there exists 
at least one hypersurface with (negative) constant mean curvature. 
Also the past direction was analyzed in areal coordinates
and global existence was established without any smallness condition
on the data. It is however not concluded if the past 
singularity in this more 
general case without the smallness condition on the data is a curvature
singularity as well. 
The question whether the areal time coordinate, which is positive by 
definition, takes all values in the range $(0,\infty)$ or only in
$(R_0,\infty)$ for some positive $R_0$ is also left open. 
In the special case in [Rn2] it is indeed
shown that $R_0=0,$ but there is an example 
for vacuum spacetimes in the more general case of 
$U(1)\times U(1)$ symmetry 
where $R_0>0.$ 

For spacetimes admitting a two-dimensional isometry group the first
study was done 
by Rendall [Rl2] in the case of local $U(1)\times U(1)$ symmetry (or
local $T^2$ symmetry). For a discussion of the topologies of 
these spacetimes we refer to the original paper. 
In the model case the spacetime is topologically of the form 
$\mathbb{R}\times T^3,$ and to simplify our discussion later on we
write down the metric in areal coordinates for this type of spacetime 
\begin{eqnarray} 
&g=\mbox{e}^{2(\eta-U)}(-\alpha dt^{2}+d\theta^{2}) 
+\mbox{e}^{-2U}t^{2}[dy+Hd\theta+Mdt]^{2}&\nonumber\\ 
&+\mbox{e}^{2U}[dx+Ady+(G+AH)d\theta+(L+AM)dt]^{2}.&\label{areal} 
\end{eqnarray} 
Here the metric coefficients $\eta,U,\alpha,A,H,L$ and $M$ depend on
$t$ and $\theta$ and $\theta,x,y\in S^1.$ In [Rl2] CMC coordinates 
are considered rather than areal coordinates. The CMC- and the areal
coordinate foliations are both geometrically based time foliations. 
The advantage with a CMC 
approach is that the definition of a CMC hypersurface does not depend
on any symmetry assumptions and it is possible that CMC foliations
will exist for rather general spacetimes. 
The areal coordinate foliation is on the other hand
adapted to the symmetry of spacetime but it has analytical advantages
which we will see below. 

Under the hypothesis that there exists at least one CMC hypersurface
Rendall proves, without any smallness condition on the data, 
that the past of the given CMC hypersurface 
can be globally foliated by CMC hypersurfaces and that the mean
curvature of these hypersurfaces blows up at the past
singularity. Again the future direction was left open. The result in
[Rl2] holds for Vlasov matter and for matter described by a wave map
(which is not a phenomenological matter model). That the choice of
matter model is important was shown by Rendall [Rl5] by a non-global 
existence result for dust, which leads to examples of spacetimes [IsR]
which are not covered by a CMC foliation. 

There are several possible subcases to the $U(1)\times U(1)$ symmetry class. 
The plane case where
the symmetry group is three-dimensional is one subcase and the form of 
the metric in areal coordinates is obtained by letting $A=G=H=L=M=0$
and $U=\log{t}/2$ in (\ref{areal}). Another subcase which still 
only admits two Killing fields (and which includes plane symmetry as a
special case) is Gowdy symmetry. It is obtained by letting $G=H=L=M=0$
in (\ref{areal}). In [An4] the author considers Gowdy symmetric spacetimes
with Vlasov matter. It is proved that the entire 
maximal globally hyperbolic spacetime can be foliated by constant
areal time slices for arbitrary (in size) initial data. The areal 
coordinates are used in a direct way for showing global existence to
the future whereas the analysis for the past direction is carried out
in conformal coordinates. These coordinates are not fixed to the
geometry of spacetime and it is not clear that the entire past has 
been covered. A chain of geometrical arguments then shows that 
areal coordinates indeed cover the entire spacetime. This method 
was applied to the problem on hyperbolic and plane symmetry in [ARR]. 
The method in [An4] was in turn inspired by the work [BCIM] for vacuum
spacetimes where the idea of using conformal coordinates in the past 
direction was introduced. As pointed out in [ARR] the result by Henkel
[He2] guarantees the existence of one CMC hypersurface in the Gowdy 
case and together with the global areal foliation in [An4] it follows that
Gowdy spacetimes with Vlasov matter can be globally covered by CMC
hypersurfaces as well (also to the future). So in a sense the areal
coordinate approach seems to be analytically favourable to the CMC
approach for these spacetimes. It remains to prove that
the general case with local $U(1)\times U(1)$ symmetry and Vlasov
matter can be 
foliated by CMC- and by constant areal time hypersurfaces. This
project is in progress by Rendall, Weaver and the author [ARW]. 
Regarding global foliations 
(with respect to a CMC- and an areal time coordinate) 
of spacetimes admitting a two-dimensional isometry 
group this result (if affirmative) will complete the picture. As
mentioned above there are however a number of important questions open 
regarding 
the nature of the initial singularity, the range of the areal coordinate
and the question of the asymptotics in the future direction. Recently,
Rein [Rn5] has shown geodesic completeness to the future for solutions 
to the Einstein-Vlasov system with hyperbolic symmetry (cf. [ARR]) 
under a certain size restriction on the initial data. 

\section{Stationary solutions to the Einstein-Vlasov system} 
Equilibrium states in galactic dynamics can be described as stationary
solutions of the Einstein-Vlasov system, or of the Vlasov-Poisson
system in the Newtonian case. Here we will consider the former case
for which only static spherically symmetric solutions have been
constructed but we mention that in the latter case also stationary
axially symmetric solutions have been found by Rein [Rn6]. 

In the static spherically symmetric case the problem can be formulated
as follows. 
Let the space-time metric have the form 
\begin{displaymath}
ds^{2}=-e^{2\mu(r)}dt^{2}+e^{2\lambda(r)}dr^{2}
+r^{2}(d\theta^{2}+\sin^{2}{\theta}d\varphi^{2}), 
\end{displaymath} 
where $r\geq 0,\,\theta\in [0,\pi],\,\varphi\in
[0,2\pi].$ As before asymptotic flatness is expressed by the boundary
conditions  
\begin{displaymath} 
\lim_{r\rightarrow\infty}\lambda(r)=\lim_{r\rightarrow\infty}\mu(r)
=0,\;\forall t\geq 0,
\end{displaymath}
and a regular centre requires 
\begin{displaymath} 
\lambda(t,0)=0. 
\end{displaymath} 
Following the notation in section 2.1 the time-independent
Einstein-Vlasov system reads 
\begin{equation} 
e^{\mu-\lambda}\frac{v}{\sqrt{1+|v|^2}}\cdot\partial_{x}f
-\sqrt{1+|v|^2}e^{\mu-\lambda}\mu_{r}
\frac{x}{r}\cdot\partial_{v}f=0,\label{Vlas3} 
\end{equation} 
\begin{eqnarray} 
&\displaystyle e^{-2\lambda}(2r\lambda_{r}-1)+1=8\pi r^2\rho,&\label{ee12}\\ 
&\displaystyle e^{-2\lambda}(2r\mu_{r}+1)-1=8\pi r^2 p.&\label{ee22} 
%&\displaystyle\lambda_{t}=-4\pi re^{\lambda+\mu}j,&\label{ee3}\\ 
%&\displaystyle e^{-2\lambda}(\mu_{rr}+(\mu_{r}-\lambda_{r})(\mu_{r}+
%\frac{1}{r}))-e^{-2\mu}(\lambda_{tt}+\lambda_{t}(\lambda_{t}-\mu_{t}))=
%4\pi q.&\label{ee4} 
\end{eqnarray} 
The matter quantities are defined as before 
\begin{eqnarray} 
\rho(x)&=&
\int_{\mathbb{R}^{3}}\sqrt{1+v^2}f(t,x,v)\;dv,\label{rho3}\\ 
p(x)&=&\int_{\mathbb{R}^{3}}\left(\frac{x\cdot
    v}{r}\right)^{2}f(t,x,v)\;\frac{dv}{\sqrt{1+v^2}}.\label{p3} 
%j(t,r)&=&\frac{\pi}{r^{2}}
%\int_{-\infty}^{\infty}\int_{0}^{\infty}wf(t,r,w,F),\;dwdF\label{j}\\ 
%q(t,r)&=&\frac{\pi}{r^{2}}\int_{-\infty}^{\infty}\int_{0}^{\infty}\frac{F}{E}f(t,r,w,F)\;
%dwdF.\label{q} 
\end{eqnarray} 
The quantities 
\begin{displaymath} 
E:=e^\mu(r)\sqrt{1+v^2},\,\,L=x^2v^2-(x\cdot v)^2=|x\times v|^2, 
\end{displaymath} 
are conserved along characteristics. $E$ is the particle energy and $L$
is the angular momentum squared. If we let $$f(x,v)=\Phi(E,L)$$ for
some function $\Phi$ the Vlasov equation is automatically
satisfied. The form of $\Phi$ is usually restricted to 
\begin{equation} 
\Phi(E,L)=\phi(E)(L-L_0)^l,\label{pol}
\end{equation} 
where $l>-1/2$ and $L_0\geq 0.$ If $\phi(E)=(E-E_0)^k_{+},\,k>-1,$ for
some 
positive constant $E_0$ this is called the polytropic 
ansatz. The case of isotropic pressure is obtained by letting $l=0$ so 
that $\Phi$ only depends on $E$. We refer to [Rn3] for information on 
the role of $L_0$. 

In passing we mention that for the Vlasov-Poisson system it
has been shown [BFH] that every static spherically symmetric solution must
have the form $f=\Phi(E,L).$ This is referred to as Jeans' 
theorem. It was an open question for some time to decide whether or
not this was 
also true for the Einstein-Vlasov system. It was settled 1999 by 
Schaeffer [Sc4] who found solutions that do not have this particular
form globally on phase space and consequently Jeans' theorem is not valid in the relativistic 
case. 

However, almost all results in this field rest on this ansatz. 
By inserting the ansatz for $f$ in the matter quantities 
$\rho$ and $p$ a nonlinear system for $\lambda$ and $\mu$ is
obtained, 
\begin{eqnarray*}
&\displaystyle e^{-2\lambda}(2r\lambda_{r}-1)+1=8\pi r^2G_{\Phi}(r,\mu),&\\ 
&\displaystyle e^{-2\lambda}(2r\mu_{r}+1)-1=8\pi r^2 H_{\Phi}(r,\mu),& 
\end{eqnarray*} 
where 
\begin{eqnarray*} 
G_{\Phi}(r,\mu)&=&
\frac{2\pi}{r^2}\int_1^{\infty}\int_0^{r^2(\epsilon^2-1)}\Phi(e^{\mu(r)}\epsilon,L)\frac{\epsilon^2}{\sqrt{\epsilon^2-1-L/r^2}}\;dLd\epsilon,\\ 
H_{\Phi}(r,\mu)&=&\frac{2\pi}{r^2}\int_1^{\infty}\int_0^{r^2(\epsilon^2-1)}\Phi(e^{\mu(r)}\epsilon,L)\sqrt{\epsilon^2-1-L/r^2}\;dLd\epsilon. 
\end{eqnarray*} 

Existence of solutions to this system was first proved in the case 
of isotropic pressure in [RR3] and then extended to the general case
in [Rn3]. 
The main problem is then to show that the resulting solutions 
have finite (ADM) mass and compact support. This is accomplished in
[RR3] for a polytropic ansatz with isotropic pressure and in [Rn3] for 
a polytropic ansatz with possible anisotropic pressure. They 
use a perturbation argument based on the fact that the Vlasov-Poisson
system is the limit of the Einstein-Vlasov system as the speed of
light tends to infinity [RR2]. Two types of solutions are constructed, 
those with a regular centre ([RR3], [Rn3]) and those ([Rn3]) with a 
Schwarzschild singularity in the centre. 
In [RR4] Rendall and Rein go beyond the polytropic ansatz and assume
that $\Phi$ satisfies 
\begin{displaymath} 
\Phi(E,L)=c(E-E_0)^k_{+}L^l+O((E_0-E)_{+}^{\delta+k})L^l\mbox{ as
  }E\rightarrow E_0, 
\end{displaymath} 
where $k>-1,\,l>-1/2,\,k+l+1/2>0,\,k<l+3/2.$ They show that this
assumption is sufficient for obtaining steady states with finite mass
and compact support. The result is obtained in a more direct way and
is not based on the perturbation argument mentioned above. Their
method is inspired by a work on stellar models by Makino [M] where he
considers steady states of the Euler-Einstein system. 
In [RR4] there is also an interesting discussion about steady states
which appear in the astrophysics literature and it is shown that 
their result applies to most of these which proves that they have
the desirable property of finite mass and compact support. 

All solutions described so far have the property that the support of
$\rho$ contains a ball about the centre. In [Rn4] Rein shows that
there exists steady states whose support is a finite, spherically
symmetric shell, so that they have a vacuum region in the centre. 

At present there are almost no known results concerning the stability
properties of the steady states to the Einstein-Vlasov system. 
In the Vlasov-Poisson case however, the nonlinear stability of
stationary solutions have been investigated by Guo and Rein [GR] using
the energy-Casimir method. In the Einstein-Vlasov case Wolansky [Wo]
has applied the energy-Casimir method and obtained some insights but
the theory in this case is much less developed than in the
Vlasov-Poisson case and the stability problem is essentially open. 

\section{Acknowledgements} 
I would like to thank Alan Rendall for helpful suggestions.

\end{document}